\documentclass[superscriptaddress,twocolumn,a4paper,showpacs,showkeys]{revtex4}
\usepackage{times}
\usepackage{epsf}
\usepackage{color}
\usepackage{graphicx}
\usepackage{amsmath,amsbsy,amsfonts,mathrsfs}
\newcommand{\be}{\begin{equation}}
\newcommand{\ee}{\end{equation}}
\newcommand{\bea}{\begin{eqnarray}}
\newcommand{\eea}{\end{eqnarray}}
\newcommand{\bg}{\begin{figure}}
\newcommand{\eg}{\end{figure}}
\newcommand{\bi}{\begin{itemize}}
\newcommand{\ei}{\end{itemize}}
\begin{document}
%%%%%%%%%%%%%%%%
%%%%%%%%%%%%%%%%

% Authors and vinculations...

%%%%%%%%%%%%%%%%%%%%%%%%%%%%%%%%%%%%%%%%%%%%%%%%%%%%%%%%%%%%%%%%%%%%%%%%%%%%%%%%%%%%%%%%%
%%%%%%%%%%%%%%%%%%%%%%%%%%%%%%%%%%%%%%%%%%%%%%%%%%%%%%%%%%%%%%%%%%%%%%%%%%%%%%%%%%%%%%%%%
\title{Optical Absorption Spectra of Ag$_{11}$ Isomers by First-Principles \\
Theoretical Spectroscopy with Time-dependent Density Functional Theory}
%%%%%%%%%%%%%%%%%%%%%%%%%%%%%%%%%%%%%%%%%%%%%%%%%%%%%%%%%%%%%%%%%%%%%%%%%%%%%%%%%%%%%%%%%
%%%%%%%%%%%%%%%%%%%%%%%%%%%%%%%%%%%%%%%%%%%%%%%%%%%%%%%%%%%%%%%%%%%%%%%%%%%%%%%%%%%%%%%%%

%%%%%%%%%%%%%%%%%%%%%%%%%%%%%%%%%%%%%%%%%%%%%%%%%%%%%%%%%%%%%%%%%%%%%%%%%%%%%%%%%%%%%%%%%
\author{\bf Jos\'e~I.~Mart\1nez}
\email{jimartinez@fysik.dtu.dk}
%\affiliation{Departamento de F\1sica Te\'orica, At\'omica y \'Optica, \\
%University of Valladolid, E-47001 Valladolid (Spain)}
\affiliation{Center for Atomic-scale Materials Design, Department of Physics NanoDTU,
Technical University of Denmark, DK-2800 Lyngby (Denmark)}
%%%%%%%%%%%%%%%%%%%%%%%%%%%%%%%%%%%%%%%%%%%%%%%%%%%%%%%%%%%%%%%%%%%%%%%%%%%%%%%%%%%%%%%%%
\author{\bf Eva~M.~Fern\'andez}
\email{efernand@fysik.dtu.dk}
\affiliation{Center for Atomic-scale Materials Design, Department of Physics NanoDTU,
Technical University of Denmark, DK-2800 Lyngby (Denmark)}
%%%%%%%%%%%%%%%%%%%%%%%%%%%%%%%%%%%%%%%%%%%%%%%%%%%%%%%%%%%%%%%%%%%%%%%%%%%%%%%%%%%%%%%%%

%%%%%%%%%%%%%%%%%%%%%%%%%%%%%%%%%%%%%%%%%%%%%%%%%%%%%%%%%%%%%%%%%%%%%%%%%%%%%%%%%%%%%%%%%%%%%
\date{\today}
%%%%%%%%%%%%%%%%%%%%%%%%%%%%%%%%%%%%%%%%%%%%%%%%%%%%%%%%%%%%%%%%%%%%%%%%%%%%%%%%%%%%%%%%%%%%%

%%%%%%%%%%%%%%%%%%%%%%%%%%%%%%%%%%%%%%%%%%%%%%%%%%%%%%%%%%%%%%%%%%%%%%%%%%%%%%%%%%%%%%%%%%%%%

% Abstract...

%%%%%%%%%%%%%%%%%%%%%%%%%%%%%%%%%%%%%%%%%%%%%%%%%%%%%%%%%%%%%%%%%%%%%%%%%%%%%%%%%%%%%%%%%%%%%%

\begin{abstract}

The optical absorption spectrum of the three most stable isomers of the
Ag$_{11}$ system was calculated u\-sing the time-dependent density functional theory,
with the generalized gradient approximation for the exchange and correlation potential, and a
relativistic pseudopotential parametrization for the modelling of the ion--electron interaction.
The computational scheme is based on a real space code, where the photoabsorption spectrum is
calculated by using the formalism develo\-ped by Casida. The significantly different spectra
of the three isomers permit the identification of the ground-state configuration predominantly
present in the laboratory beams in base to a comparison between the calculated photoabsorption
spectrum of the most stable configuration of Ag$_{11}$ and the measured spectra of medium-size
silver clusters trapped in noble gas Ar and Ne matrices at different temperatures.
This assignment is confirmed by the fact that this isomer has the lowest calculated energy.

\end{abstract}
%%%%%%%%%%%%%%%%%%%%%%%%%%%%%%%%%%%%%%%%%%%%%%%%%%%%%%%%%%%%%%%%%%%%%%%%%%%%%%%%%%%%%%%%%%%%%%

% Pacs and Keywords...

%%%%%%%%%%%%%%%%%%%%%%%%%%%%%%%%%%%%%%%%%%%%%%%%%%%%%%%%%%%%%%%%%%%%%%%%%%%%%%%%%%%%%%%%%%%%%%


%%%%%%%%%%%%%%%%%%%%%%%%%%%%%%%%%%%%%%%%%%%%%%%%%%%%%%%%%%%%%%%%%%%%%%%%%%%%%%%%%%%%%%%%%%%%%%

\maketitle

%%%%%%%%%%%%%%%%%%%%%%%%%%%%%%%%%%%%%%%%%%%%%%%%%%%%%%%%%%%%%%%%%%%%%%%%%%%%%%%%%%%%%%%%%%%%%%

% Document...

%%%%%%%%%%%%%%%%%%%%%%%%%%%%%%%%%%%%%%%%%%%%%%%%%%%%%%%%%%%%%%%%%%%%%%%%%%%%%%%%%%%%%%%%

%%%%%%%%%%%%%%%%%%%%%%
\section{INTRODUCTION}
%%%%%%%%%%%%%%%%%%%%%%

The study of transition and noble metal clusters has focused considerable
attention along the last years due to the diffe\-rent properties detected respect
to their corresponding bulk-phases. Additionally, these cluster properties have
de\-monstrated to be strongly dependent on the cluster size. On the other hand,
noble metal clusters, and silver clusters in particular, present important
applications from the scientific and technological point of view in interesting
researching fields such as catalysis~\cite{pykko04,fernandez05,zhou07,fernandez06,qu04,zhou06},
and optics~\cite{peyser01,lee02,gleitsmann04,peyser05}.

The structural and electronic properties of silver
clusters~\cite{vlasta01,FernandezPRB04,Fournier,huda03,lee03,yang06}
have been widely studied, however unlike the optical properties~\cite
{vlasta01,yabana99,Idrobo,harbich92,Fedrigo,Conus,Lee,Harbich}.
The optical absorption spectra of silver clusters have been studied for
more than two decades ago. Fedrigo {\it et al.}~\cite{harbich92,Fedrigo}
found that Ag$_N$ (N$\leq$40) clusters show a simple dominant peak in the
low energy region, exceptions were attributed to the coexistence of low energy
isomers. These results were confirmed theoretically for small-sized silver
clusters (n=5-8)~\cite{vlasta01}. On the other hand, Idrobo {\it et al.}~\cite{Idrobo}
have recently carried out calculations of the absorption spectra of Ag$_{11}$ for
different structural isomers, however no full agreement with experiments was found.

On the other hand, experimental optical spectroscopy, aided by recent
advances in the corresponding theoretical tools, has proved to be a
powerful means for obtaining information about the geometrical and
electronic structure of molecules and small clusters. Since the structure
of a cluster is gene\-rally not directly accessible experimentally, its
characterization must be done with a combination of experiment and theory.
In previous works we have shown how a comparison of the measured optical
spectrum with theoretical calculations for different isomeric forms of a given
cluster can provide a useful diagnosis of the geometrical structure of the
cluster~\cite{Martinez_TMC2,MartinezPRB07}. However, conventional optical
absorption spectroscopy experiments for studying transition and noble metal
clusters are difficult due to several technical issues, such as the extremely
low density of the produced clusters, and theoretical spectroscopy is a perfect
framework to complement most of experimental analysis.

In this paper, we present calculations of the optical
absorption spectra of the three most stable isomeric forms
of the Ag$_{11}$ cluster up to 8 eV by using an extremely accurate
technique for treating the electronic excitations: the time-dependent
density functional theory (TDDFT)~\cite{tddft1,tddft2,Castro04} under
the forma\-lism develo\-ped by Casida {\it et al.}~\cite{Casida,Jamorski96}.

\begin{figure}[t]
\centerline{\includegraphics[width=6.5cm]{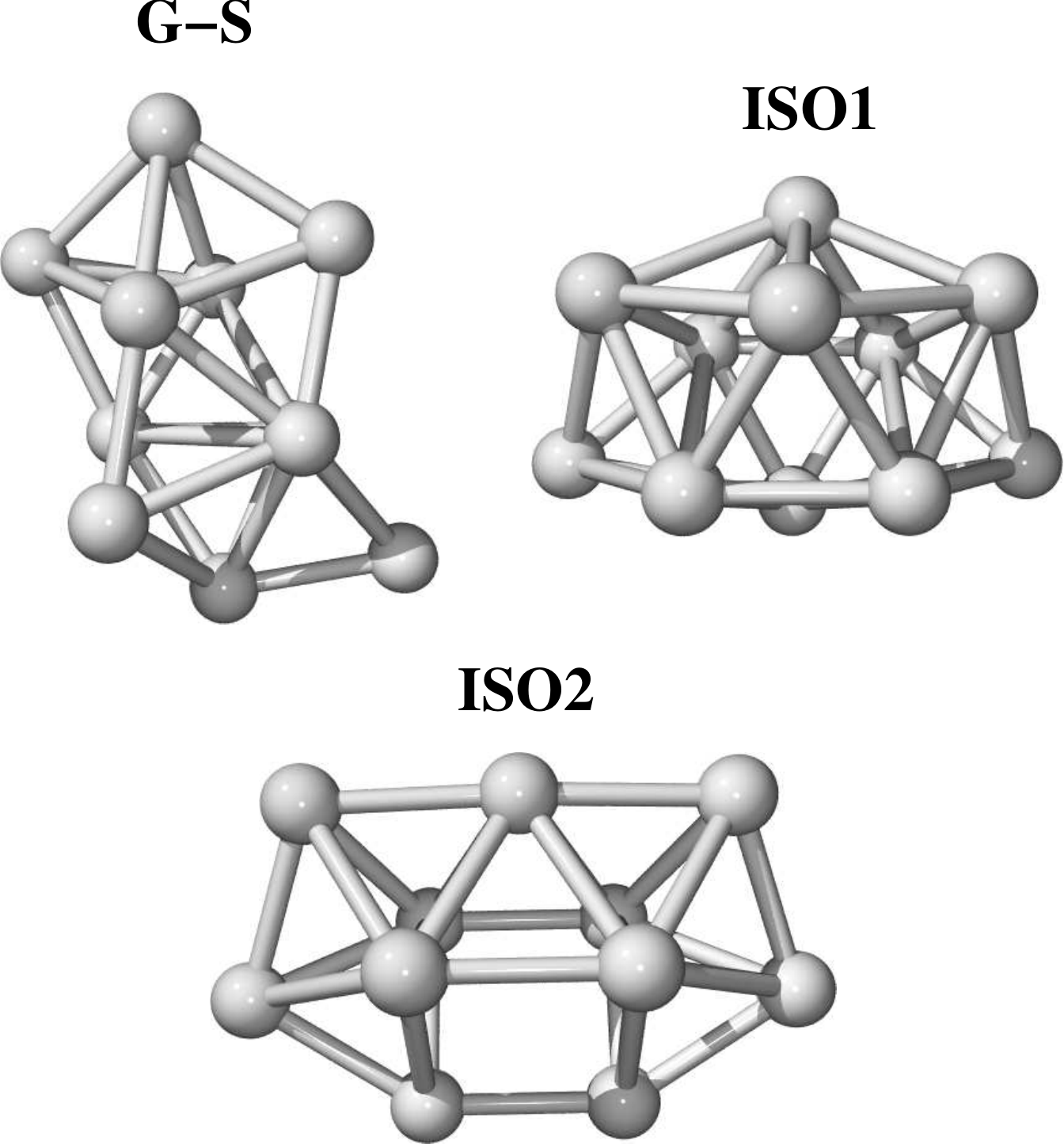}}
\smallskip \caption{View of the three most stable structures of
Ag$_{11}$ used for the calculation of the photoabsorption spectra.
The ground-state structure is labelled as G-S, and the two closest
in energy isomers as ISO1 and ISO2, respectively.
\label{P1}}
\end{figure}

Our motivations for this work have been the following:
(a) first, to show a comparison between the present first-principles
TDDFT results and the available experimental measurements, in order to
improve the description of previous theoretical studies by reducing the
numerical uncertainty; (b) since signi\-ficant differences have been found among
the spectra of the different isomers of Ag$_{11}$ (see Fig.~\ref{P1}), to propose
the compa\-rison of experimental and TDDFT absorption spectra as a power\-ful tool to elucidate
between the different isomeric forms of small clusters; and (c) finally, to predict
Ag$_{11}$ photoabsorption spectrum for energies higher than 5.5 eV, a range of energies
which has not been considered in previous experi\-mental or theoretical studies for this
system (our calculations extend the range of excitation energies up to 8 eV).

%%%%%%%%%%%%%%%%%%%%%%%%%%%%%%%%%%%%%%%%%%%%%%%%%%%%%%%%%%%%%%%%%%
%%%%%%%%%%%%%%%%%%%%%%%%%%%%%%%%%%%%%%%%%%%%%%%%%%%%%%%%%%%%%%%%%%

%%%%%%%%%%%%%%%%%
\section{METHOD}
%%%%%%%%%%%%%%%%%

%%%%%%%%%%%%%%%%%%%%%%%%%%%%%%%%%%%%%%%%%%%%%%%%%%%%%%%%%%%%%%%%%%
\begin{figure*}
\centerline{\includegraphics[width=15cm]{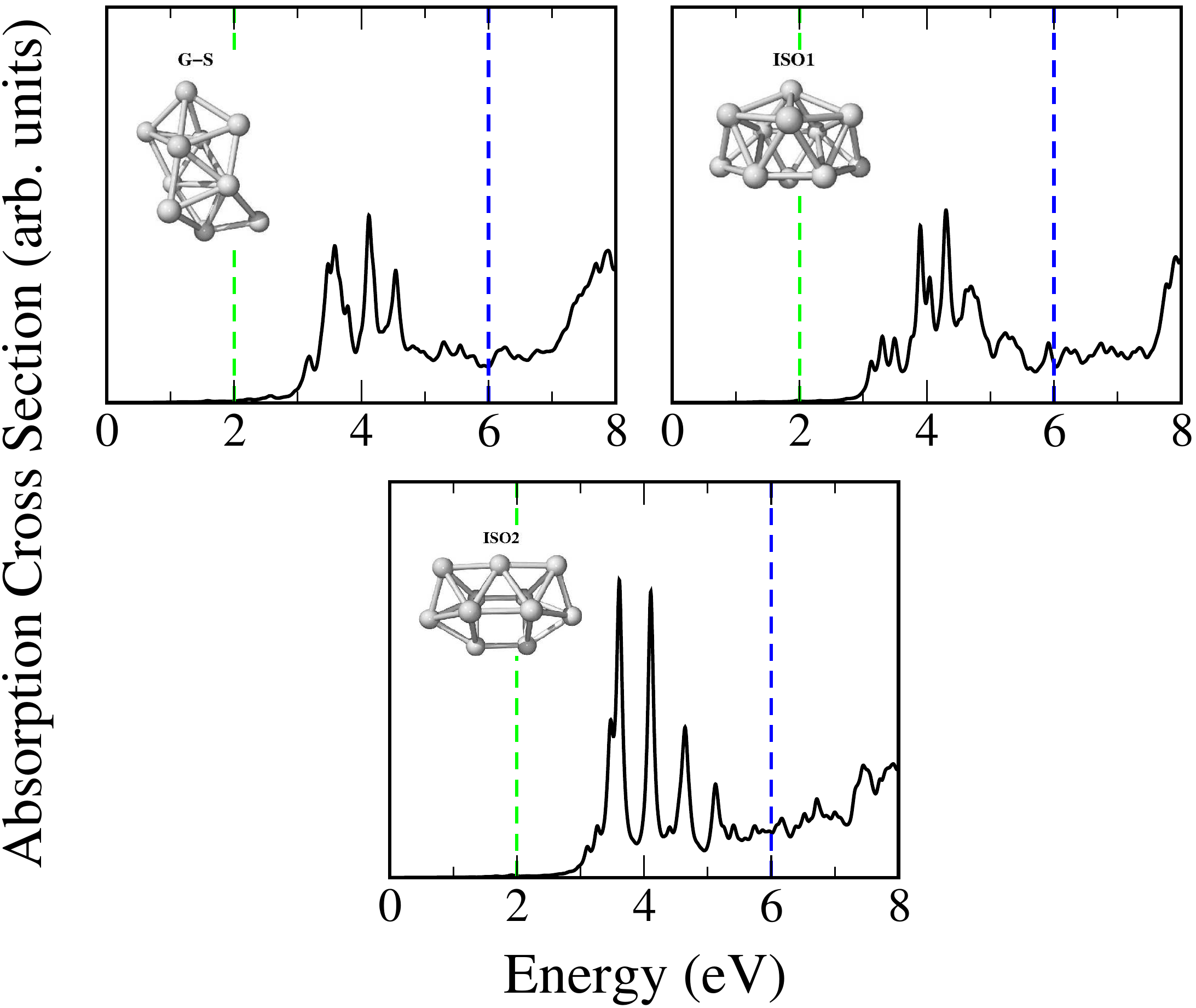}}
\smallskip \caption{Calculated photoabsorption cross sections
(in arbitrary units) for the three most stable isomers of Ag$_{11}$
(see Fig.~\ref{P1}) for energies up to 8 eV. All the absorption
strengths are shown in the same scale, and superposed to the
photoabsorption profiles vertical dashed lines have been included
for a better visualization of the different regions of the spectra.
\label{P2}}
\end{figure*}
%%%%%%%%%%%%%%%%%%%%%%%%%%%%%%%%%%%%%%%%%%%%%%%%%%%%%%%%%%%%%%%%%%

Prior to the calculation of the photoabsorption spectra, the lowest
energy geometrical structure of each isomer was determined by using
the Density Functional Theory~\cite{Kohn65}. The ion-electron interac\-tion is mo\-delled by
repla\-cing the ionic cores (1s$^2$2s$^2$p$^6$3s$^2$p$^6$d$^{10}$4s$^2$p$^{6}$d$^{10}$5s$^1$
Krypton--like core) by pseudopotentials within the relativistic scheme of
Hartwigsen and coworkers~\cite{HGH}, pre\-viously validated for seve\-ral
metal elements in pure and organometallic clusters~\cite{Martinez_TMC1,Martinez_TMC2,MartinezPRB07}.
This pseudopotential is built in such a way that is sepa\-rable by cons\-truction, is
highly accurate, and has an ana\-lytical form that can be specified
by a small number of para\-meters. For exchange and corre\-lation
effects we have used the gene\-ralized gradient approximation
(GGA)~\cite{PBE96}.

As initial geometries for the structural optimization of the
Ag$_{11}$ isomers were considered the three most stable structures previously obtained by
Fern\'andez \textit{et al.}~\cite{FernandezPRB04}. Starting from those
geometries we performed structural relaxations by making use of the
Broyden algorithm~\cite{Broyden} with a convergence criterion of
$10^{-2}$~eV/\AA~in the net forces on every atom. The structural
relaxations, performed with the {\sc abinit} packa\-ge~\cite{abinit},
did not produce significant variations with res\-pect to the ori\-ginal
structures. In order to test the influence of using a well-tailed
exchange-correlation potential in the structural properties we have
additio\-nally checked the simple GGA by compari\-son with a functio\-nal
with improved asymptotic behavior: the van Leeuwen--Baerends (LB94)
GGA functio\-nal~\cite{LB94}, which led to practically identical
optimized structures. The geometries obtained, which are then used in the
calculations of the photoabsorption spectra, are shown in Fig.~\ref{P1}.

Once the ground-state structures of the different isomers were
established, we performed the calculation of the excitation
spectra. For this purpose, we have used the TDDFT formalism~\cite{tddft1,tddft2,Castro04},
implemented in the real space code {\tt octopus}~\cite{octopus}. The
theoretical foundations underlying the TDDFT calculations, as well
as the computational scheme, have been presented
elsewhere~\cite{octopus,Petersilka96,Castro2004,Casida}, and here
we only summarize the main points. The photoabsorption spectrum is
calculated by using the formalism develo\-ped by
Casida~\cite{Casida,Jamorski96}. Application of this technique to
obtain the oscillator strengths requires a pre\-vious calculation of
the ground state electronic structure, that is, the occupied
electronic states, and also the unoccupied states. The whole set of
occupied states, joint to twenty extra unoccupied states for each spin
channel, were ne\-cessary to reach convergence of the excitation
spectra for an energy range up to 8 eV. After that, each excitation peak is broa\-dened by a Lorentzian
profile to give the photoabsorption cross section
\begin{equation} \label{PS}
\sigma_{\textrm{abs}}(\epsilon)=\sum_{\epsilon_{i}}\frac{A^{2}}{(\epsilon-\epsilon_{i})^{2}+A^{2}},
\end{equation}
where $\epsilon$ is the energy, $\epsilon_{i}$ are the
discrete excitation energies obtained by the Casida method, and the
value of the parameter $A$ determines the full width at half maximum.

We have performed the calculations of the photoabsorption spectra
with the GGA, using the Perdew-Burke-Ernzerhof parame\-terization~\cite{Ernzerhof} of
electronic corre\-lation. The corres\-ponding occupied and unoccupied single--particle states
are used in the Casida method with the adiabatic LDA exchange--correlation
kernel. Although not entirely consistent, this is a common practice because
the use of a better kernel does not significantly affect the results as
long as the single--particle orbitals have been calculated with a
suitable static $v_{xc}({\textbf r})$ potential~\cite{MartinezPRB07}. This computational framework
has been used successfully in the calculation of the optical spectra of atoms and clusters~\cite{Martinez_TMC1,Martinez_TMC2,MartinezPRB07,Martinez_MC,Rubio96,Castro2002,Vasiliev,rubio97,Marques01,Tsolakidis,Koponen}.
In the case of free atoms, the energies for one-particle low-energy excitations agree with expe\-riment
to within 5-10 per cent~\cite{Vasiliev}. For clusters of s--p metals, the errors in the position of the
absorption peaks are usually in the 0.1-0.2 eV range~\cite{rubio97,Vasiliev,Marques01}, and the results are also
very reaso\-nable for carbon and BN clusters~\cite{Tsolakidis,Koponen}, and transition metal, organometallic, and molecular and van der Waals clusters~\cite{Martinez_TMC1,Martinez_TMC2,MartinezPRB07,Martinez_MC}, even in the high-energy range.

This formalism, used in combination with both adiabatic local density (ALDA) and GGA
approximations for the XC effects, has provided successful results, establishing this technique
for the calculation of excitations in this kind of nanoscale systems as an effective
tool in the characterization of a great number of cluster properties and behaviors,
such as the elucidation of ground-state structures, the characterization
of dimeric growths in molecular clusters, and the theoretical justification of collective
electronic behaviors~\cite{Martinez_TMC1,Martinez_TMC2,MartinezPRB07,Martinez_MC}.

%%%%%%%%%%%%%%%%%%%%%%%%%%%%%%%%%%%%%%%%%%%%%%%%%%%%%%%%%%%%%%%%%%%%%%%%%%%%%%%%%%%%%%%
\section{Results and discussion}
%%%%%%%%%%%%%%%%%%%%%%%%%%%%%%%%%%%%%%%%%%%%%%%%%%%%%%%%%%%%%%%%%%%%%%%%%%%%%%%%%%%%%%%

%%%%%%%%%%%%%%%%%%%%%%%%%%%%%%%%%%%%%%%%%%%%%%%%%%%%%%%%%%%%%%%%%%
\begin{figure}
\centerline{\includegraphics[width=\columnwidth]{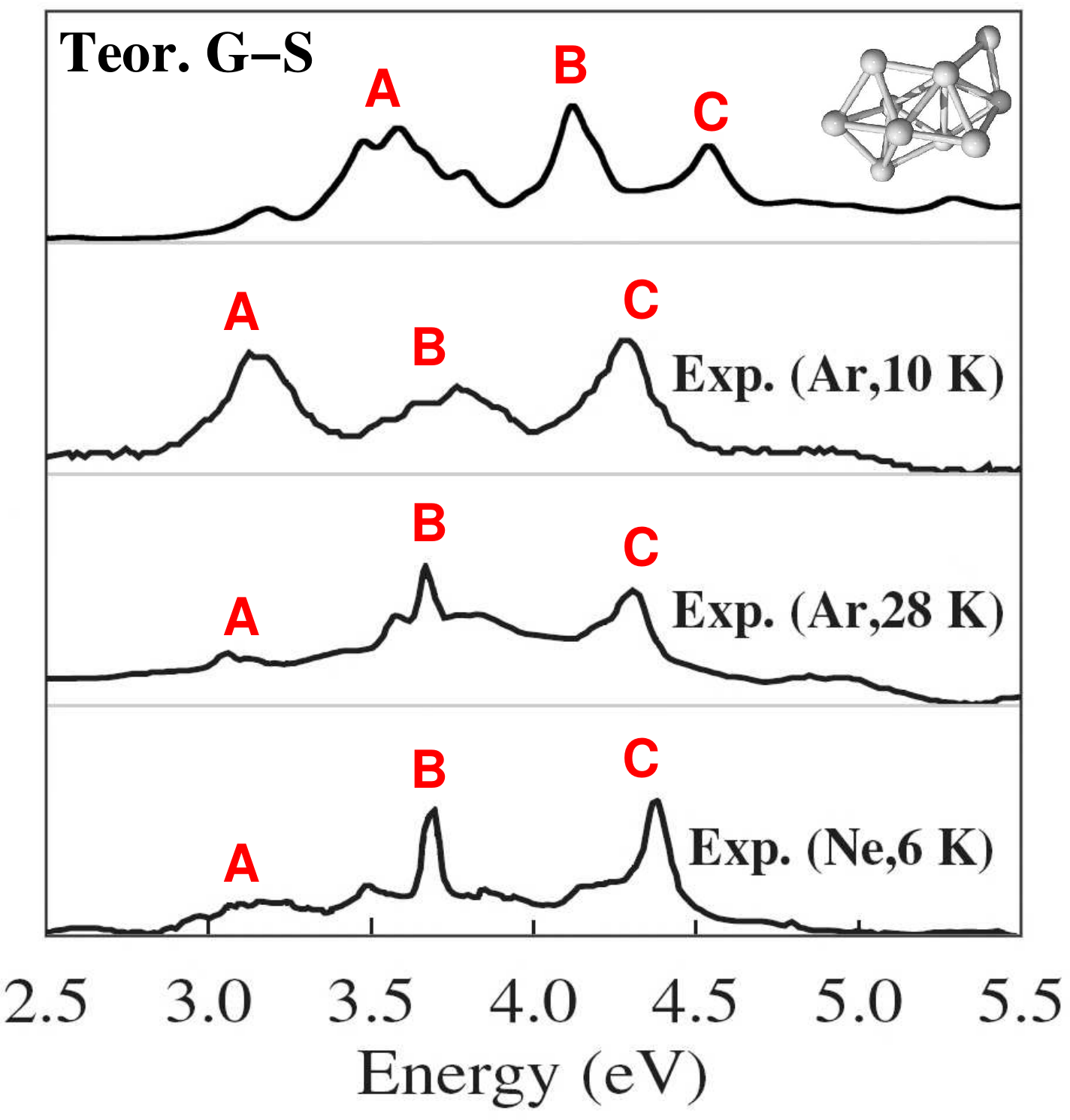}}
\smallskip \caption{Calculated optical absorption spectrum of the
G-S structure of Ag$_{11}$ (upper pannel), and three experimental
spectra of Ag$_{11}$ measured in a Ne matrix at 6 K~\cite{Harbich},
and Ar matrices at 10~\cite{Fedrigo} and 28 K~\cite{Conus}
(lower pannels). The computed spectrum has been broadened using
Lorentzian profiles with a width of 0.05 eV.
\label{P3}}
\end{figure}
%%%%%%%%%%%%%%%%%%%%%%%%%%%%%%%%%%%%%%%%%%%%%%%%%%%%%%%%%%%%%%%%%%

Fig.~\ref{P1} shows the calculated equilibrium structures of the three
lowest-energy isomers of the Ag$_{11}$ system, starting from the geometries
obtained by Fern\'andez \textit{et al.}~\cite{FernandezPRB04}. The
results do not reveal significant variations respect the original
structures; where, during the relaxation process, all the atoms in the
clusters are free to move. The obtained G-S configuration is based in a seven-atom
pentagonal bypiramid, as predicted in previous references~\cite{Idrobo,Fournier,Lee}.
The ordering of the most stable isomers presented here do not fully agree with these
previous studies, but is in complete agreement with the accurate structural analysis given by
Fern\'andez \emph{et al.} (for a more detailed description about the structures see
Ref.~\cite{FernandezPRB04}). However, this factor can be due to the small differences in
the total energy detected between the different, almost energy-degenerated,
isomers shown in the recent lite\-rature (in the range of meV). In the present
study the gap in the total energy between the G-S structure and the two most
stable isomers is 0.08 and 0.15 eV for ISO1 and ISO2 structures, respectively.
Nevertheless, and according to the previous theo\-retical studies, the ground-state
of this kind of systems has demonstrated to be highly sensitive to the common DFT
numerical implementations.

Figure~\ref{P2} shows the photoabsorption cross sections for the
three most stable isomers of Ag$_{11}$ up to 8 eV as obtained by
TDDFT under the Casida formalism. The finite width of the
absorption peaks in an experiment -- linked to the accessible
resolution -- is mostly determined by the temperature. In our
calculations, on the other hand, the width of the peaks is an
artifact and may be reduced or enlarged by modifying the $A$
parameter. We have broadened the peaks in the way shown in the
figure (with $A$=0.05 eV), so as to simulate the approximate
resolution yielded by most of the experiments of photoabsorption
and photodissociative spectroscopy~\cite{Martinez_TMC2}.

All three spectra are rich in features in the whole energy region
up to 8~eV from the begining of the absorption. Three regions can
be clearly distinguished in all the spectra (separated by dashed
lines superposed to each profile in the picture). A region of no
absorption at all covers the energy range up to 2 eV (common to all
isomers), a region of strong absorption is observed at energies between
2 and 6 eV, and another region of weaker absorption at energies larger
than 6 eV, with an absorption strength dependent on the isomer in particular.
To our knowledge, no experimental or theoretical study has been carried out
for Ag$_{11}$ in a range of excitation energies up to 8 eV. In the region
between 2 and 6 eV, where the absorption strength is more pronounced,
the spectra are dominated by a sharped profile whose position rather
varies between the different clusters, and formed by pronounced peaks
at 3.6, 4.2 and 4.6 eV. For the ISO1 and ISO2 isomers an additional peak
at 5.2 eV is found, not present for the G-S structure. Although this wide
absorption region is always located between 2 and 6 eV, and despite of other
previous TDDFT theoretical photoabsorption studies in metallic clusters
~\cite{Martinez_TMC1}, it cannot be interpreted as a collective plasmon since
it is possible to differenciate each single-particle excitation forming the
profile. It is noticeable that the most pronounced peaks forming this strong
absorption region between 2 and 6 eV are given for the ISO2 isomer, due to the
dege\-neracy of intermediate electronic states arising from its high
symmetry. In fact, the spectra of the other two isomers, where
the electronic degeneracy of the intermediate occupied states
is much lower, do not present such distinct absorption peaks.
On the other hand, in the high excitation energy region (6-8 eV) all the isomers show
a first zone of moderate absorption up to 7 eV, which is not dependent on the
particular geometry, and a final part where the strength of the absorption
increases. As in previous papers~\cite{Martinez_TMC1,MartinezPRB07}, we
propose that at this energy range the high excitation energies can be
interpreted for this system in terms of the contributions between
transitions involving strongly hybridated $s-d$ electronic orbitals.
It is noticeable that this effect is less important for the ISO2 structure,
where, according to its high symmetry and Kohn--Sham levels diagram, the
$sd$ hybridation is rather pronounced.

The low energy response of these isomers up to 5.5 eV
can be compared with the experimental results of the Refs.
~\cite{Fedrigo},~\cite{Conus}, and~\cite{Harbich} for the
measured spectra of medium-size silver clusters trapped in noble
gas matrices at several tempe\-ratures. Figure~\ref{P3} shows the
experimental absorption cross sections of Ag$_{11}$ in a Ne
matrix at 6 K, and Ar matrices at 10 and 28 K (lower pannels),
and the TDDFT spectrum for the obtained Ag$_{11}$ G-S structure
(upper pannel). The estimated uncertainties of the relative cross
sections in this kind of experiments use to be around 10\% (the
uncertainties in the absolute cross sections are even larger).
The experimental absorption cross sections show a clear three
peaks structure, where the relative strength of the peaks depend
strongly on the tempe\-rature and the matrix surrounding the trapped
cluster, but not their excitation energies. The peaks arise at 3.2, 3.7
and 4.4 eV, with a region of no absorption up to 5.5 eV. The TDDFT
spectrum, for the G-S isomer, shows a similar three peaks profile located at 3.6, 4.2 and 4.6 eV.
As we can appreciate in Fig.~\ref{P3}, there exist a very good agreement between
the calculated and the experimental peak positions if we just allow for a constant
shift of around 0.2-0.4 eV. This shift to the red of the experimental spectra
can be justified by the additional effect of the nobel gas matrices to the photoabsorption
response, which has been quantified by Fedrigo \emph{et al.}~\cite{Fedrigo}, and Sieber
\emph{et al.}~\cite{Sieber} in 0.2-0.25 eV. However, most of authors coincide
in that value of the shift may not be appropriate because additional contributions
could arise, such as compression effects of the cluster by the matrix, or
matrix-induced polarization of the cluster charge density~\cite{Idrobo}.
On the other hand, previous authors suggest that the deviation in the experimental
photoabsorption spectra of the small silver clusters showing several isolated peaks,
instead of only one pronounced excitation detected in large-sized clusters~\cite{harbich92,Fedrigo},
is due to the presence of se\-veral isomers in the experimental beams~\cite{Idrobo,Conus}.
However, the present study reveal theoretical spectra for three different Ag$_{11}$ isomers where such
isolated peaks are found. From these results, we can conclude that the presence of these distinct
peaks in experimental spectra is not only due to the existence of several isomeric
forms in the experimental samples since it seems to be an intrinsic optical property
of this kind of aggregates. Other discrepancy sources in the spectrum may be due to
additional factors:
(i) the error introduced in the calculation by the necessarily
approximated exchange and correlation potential,
(ii) the expectable influence of the silver clusters
coupling to the surrounding noble metal matrices, and
(iii) the possible presence in the experimental sample of
a mixture of various isomers, coinciding with other authors.

The calculated spectrum for this isomer provides the best matching with experiment,
and this confirms the assignment of the ground state structure as predicted
even by the total energies. We then conclude that the calculated spectrum of the
most stable isomer obtained provides a very accurate description of
the experimental cross section and permits to conjecture with
a high confidence that the isomer present most abundantly in the Ag$_{11}$
cluster beams has that structure. This is reasonable, since this
is indeed the calculated lowest energy isomer.

%%%%%%%%%%%%%%%%%%%%%%%%%%%%%%%%%%%%%%%%%%%%%%%%%%%%%%%%%%%%%%%%%%%%%%%%%%%%%%%%%%%%%%%
\section{Summary and conclusions}
%%%%%%%%%%%%%%%%%%%%%%%%%%%%%%%%%%%%%%%%%%%%%%%%%%%%%%%%%%%%%%%%%%%%%%%%%%%%%%%%%%%%%%%

We have performed calculations of the optical absorption
cross sections of three structural isomers of Ag$_{11}$.
For that purpose we have first employed DFT to optimize
the geometries of those isomers. For each of those structures
TDDFT has been used to calculate the photoabsorption
spectrum. In the calculations, we used the generalized
gradient approximation for the exchange and correlation
potential, relativistic norm-conserving pseudopotentials,
and the Casida forma\-lism to calculate the excitation spectrum.
This approach yields spectra for the different isomers that
are significantly diffe\-rent. The spectrum corresponding to the
most stable structure obtained matches well the experimental
results, and we conclude that this is the isomer predominantly
present in the labo\-ratory cluster beams; confirmation comes from the
fact that this is the isomer with the lowest calculated energy.
The efficiency of DFT and TDDFT for the calculation of
both ground state and spectroscopic features of molecules
and clusters permits to use these theoretical methods, as
we have shown here, as a complement to the experimental
analysis, which in many occasions lacks some important
information, such as the precise geometrical atomic
arrangement of the probed systems.

%%%%%%%%%%%%%%%%%%%%%%%%%%%%%%%%%%%%%%%%%%%%%%%%%%%%%%%%%%%%%%%%%%%%%%%%%%%%%%%%%%%%%%%
\section{ACKNOWLEDGMENTS}
%%%%%%%%%%%%%%%%%%%%%%%%%%%%%%%%%%%%%%%%%%%%%%%%%%%%%%%%%%%%%%%%%%%%%%%%%%%%%%%%%%%%%%%

The Center for Atomic-scale Mate\-rials Design is funded by the Lundbeck Foundation.
The authors wish to acknowledge additional support from the Danish Research Agency
through Grant 26-04-0047 and the Danish Center for Scientific Computing through Grant
HDW-0107-07. J.~I.~M and E.~M.~F. wish also to acknowledge the interes\-ting scientific
discussions with Professors J.~A.~Alonso and L.~C.~Balb\'as.

%%%%%%%%%%%%%%%%%%%%%%%%%%%%%%%%%%%%%%%%%%%%%%%%%%%%%%%%%%%%%%%%%%%%%%%%%%%%%%%%%%%%%%%

%%%%%%%%%%%%%%%%%%%%%%%%%%%%%%%%%%%%%%

%%%%%%%%%%%%%%%%%%%%%%

%%%%%%%%%%%%%%
%%%%%%%%%%%%%%
\end{document}